\documentclass[prd,twocolumn, superscriptaddress, nofootinbib, nobalancelastpage preprintnumbers]{revtex4-1}

\usepackage[bottom]{footmisc}
\usepackage{hyperref}
\usepackage[normalem]{ulem}
\usepackage[utf8]{inputenc}
\usepackage{amsmath}
\usepackage{graphicx}
\usepackage{braket}
\usepackage{color}
\usepackage[utf8]{inputenc}

\newcommand{\lb}{\left (}
\newcommand{\rb}{\right )}

\newcommand{\ob}{\Omega_{b}}
\newcommand{\om}{\Omega_{m}}
\newcommand{\oL}{\Omega_{\Lambda}}
\newcommand{\dl}{\delta_L}

\newcommand{\msun}{\mathrm{M}_{\odot}}
\newcommand{\lcdm}{\Lambda\mathrm{CDM}}
\newcommand\lsim{\mathrel{\rlap{\lower4pt\hbox{\hskip1pt$\sim$}}\raise1pt\hbox{$<$}}}
\newcommand\gsim{\mathrel{\rlap{\lower4pt\hbox{\hskip1pt$\sim$}}\raise1pt\hbox{$>$}}}
\newcommand{\graphic}[2]{\includegraphics[width=#2\linewidth, type=pdf,ext=.pdf,read=.pdf]{#1}}

\begin{document}
	
	\preprint{YITP-SB-19-26}
	\title{Separate Universe Void Bias}
	\author{Drew Jamieson and Marilena Loverde \\{\it{\small  C.N. Yang Institute for Theoretical Physics, Department of Physics \& Astronomy, Stony Brook University, Stony Brook, NY 11794}}}
	
	\begin{abstract}
		Voids have emerged as a novel probe of cosmology and large-scale structure. These regions of extreme underdensity are sensitive to physics beyond the standard model of cosmology, and can potentially be used as a testing ground to constrain new physics. We present the first determination of the linear void bias measured in separate universe simulations. Our methods are validated by comparing the separate universe response bias with the clustering bias of voids. We find excellent agreement between the two methods for voids identified in the halo field and the down-sampled dark matter field. For voids traced by halos, we identify two different contributions to the bias. The first is due to the bias of the underlying halo field used to identify voids, while the second we attribute to the dynamical impact of long-wavelength density perturbations on void formation and expansion. By measuring these contributions individually, we demonstrate that their sum is consistent with the total void bias. We also measure the void profiles in our simulations, and determine their separate universe response. These can be interpreted as the sensitivity of the profiles to the background density of the Universe.
	\end{abstract}

	\maketitle

	\section{Introduction}
		\label{sec:intro}

		Underdensities in the initial density field of the Universe expand to form cosmic voids in late-time, large-scale structure. In recent years, voids have emerged as a competitive source of information about our Universe \cite{Sutter:2012tf, Sutter:2014oca, Kitaura:2015ubm, Hamaus:2016wka, Mao:2016faj, Nadathur:2017jos}. The dynamics and statistics of voids provide information about the expansion history, matter contents, and initial conditions of our Universe. 

		While there are a number of challenges to using voids to constrain new physics, such as their low number density, as well as potential dependence on the precise definition of what constitutes a void (e.g. \cite{Padilla:2005ea, Colberg:2008qg, Jennings:2013nsa, Nadathur:2015lha, Lares:2017qbk, Contarini:2019qwf}), there are also a number of advantages. The interiors of voids undergo a much milder evolution compared with halos, which collapse and virialize, so voids tend to preserve their initial conditions \cite{Neyrinck:2012bg, Falck:2014fra}. Voids also become dark energy dominated earlier than the mean universe and can be used to study dynamical models of dark energy  (e.g. \cite{Lee:2007kq, Lavaux2009, Biswas:2010ey, Lavaux2012, Bos:2012wq, Hamaus:2013qja, Hamaus:2015yza, Pisani:2015jha, Endo:2018xhx, Verza:2019tvg}) and modified gravity (e.g. \cite{Li:2010zzx, Li:2011pj, Clampitt:2012ub, Lam:2014kua, Hamaus:2014afa, Zivick:2014uva, Voivodic:2016kog, Nadathur:2016nqr, Falck:2017rvl, Cautun:2017tkc, Paillas:2018wxs, Perico:2019obq}). The underdense environment in voids can be used as a laboratory to study the environmental dependence of galaxy evolution (e.g. \cite{Rojas:2003hf, Rojas:2004za, vandeWeygaert:2011as}), and they are also particularly sensitive to warm dark matter and neutrinos \cite{Yang:2014upa, Massara:2015msa, Banerjee:2016zaa, Kreisch:2018var}.  While most of the matter of the Universe ends up in collapsed objects at late times, voids account for the majority of the late-time volume of the universe. For a recent summary of cosmological opportunities with voids see \cite{Pisani:2019cvo} and references therein. 

		Voids are biased tracers of the underlying density field \cite{Sheth:2003py, Hamaus:2013qja, Chan:2014qka, Clampitt:2015jra}. In order to use void clustering to constrain cosmology (e.g. \cite{Banerjee:2016zaa, Chan:2018piq}), their bias must be measured and understood. The separate universe method is an ideal tool for determining the void bias. In the separate universe approach, we study how the observables of cosmic structure are affected by the presence of long-wavelength density perturbations. Long-wavelength modes can be absorbed into a locally defined, separate universe background density. A region embedded in a long-wavelength overdensity has a separate universe expansion history corresponding to a closed Friedmann-Robertson-Walker (FRW) cosmology with a small, positive spatial curvature. A region embedded in a long-wavelength underdensity similarly evolves as an open FRW cosmology with a small, negative spatial curvature. These slight differences in expansion histories lead to differences in structure formation. The fractional change in the number density of voids between overdense and underdense separate universe expansion histories corresponds to the linear void bias.

		The separate universe formalism describes the weakly nonlinear aspects of cosmic structure formation, where long-wavelength, linear modes couple to and affect the evolution of small-scale modes in the density field. While we can compute the evolution of long-wavelength modes using perturbation theory, the small-scale physics can be simulated, for instance using N-body simulations. Separate universe simulations involve pairs of simulations that are run with different background expansion histories corresponding to a long-wavelength overdensity and underdensity. By using the same set of random initial conditions for the pairs of simulations, cosmic variance will cancel in the linear responses of the separate universe observables. This is known to lead to improvements over other methods in the determining quantities such as the linear halos bias \cite{Wagner:2014aka, Li:2015jsz, Baldauf:2015vio, Lazeyras:2015lgp}.  

		Separate universe techniques were developed for studying the production of non-Gaussianity during inflation, using calculations of the separate universe response of the power spectrum, which corresponds to the squeezed-limit of the bispectrum \cite{Maldacena:2002vr, Creminelli:2004yq, Dai:2015jaa}. The formalism is also used to study nonlinear gravitational evolution in large-scale structure. Applications of separate universe methods in N-body simulations have measured gravitationally induced non-Gaussianty, through the separate universe response of the matter power spectrum, as well as measuring the halo bias and the response of other small-scale observables \cite{McDonald:2001fe, Sirko:2005uz, Gnedin:2011kj, Wagner:2014aka, Li:2014sga, Li:2014jra, Li:2015jsz, Chiang:2014oga, Manzotti:2014wca, Baldauf:2015vio, Lazeyras:2015lgp, Paranjape:2016pbh}. The separate universe has also been used to study scale-dependent halo bias and scale-dependent power spectrum responses in  cosmologies beyond $\Lambda$CDM \cite{Hu:2016ssz}, including dynamical dark energy with adiabatic initial conditions \cite{Chiang:2016vxa}, massive neutrinos \cite{Chiang:2017vuk}, dynamical dark energy with isocurvature perturbations \cite{Jamieson:2018biz}, and also with baryon-CDM isocurvature \cite{Barreira:2019qdl}. In this work, we present the first application of separate universe simulations to measure void bias. We also measure the sensitivity of void profiles to the background matter density. 

		This paper is organized into the following sections. Section \ref{sec:su} reviews the separate universe formalism. In Section \ref{sec:sim}, we present the set up and details of our N-body simulations, and explain our analysis of the simulation snapshots. Here we also define the linear response observables and explain the changes that were made to the N-body simulations, and the halo- and void-finding algorithms. The main results for the linear void bias of voids found in halos are presented in Section \ref{sec:bias}. We also present our decomposition of the void bias in this section, and show the comparison between the separate universe response bias and the clustering bias. The void profiles, along with their separate universe response, are presented in Section \ref{sec:prof}. While the voids found in halos are more closely related to voids found in observations of galaxy catalogs, voids found in dark matter are often used as the basis for model building.  We include results for the void bias, and void profile responses of voids traced by dark matter in Section \ref{sec:dm}. Our conclusions are presented in Section \ref{sec:con}.

	\section{Separate Universe Formalism}
		\label{sec:su}
	
		A density perturbation that is sufficiently large in scale appears effectively homogenous from the point of view of small-scale observables. These small-scale observables, in the context of a long-wavelength density perturbation, evolve as if they belong to a separate universe in which the mean density of the background cosmology is shifted by the long-wavelength mode. To make this concrete, we denote observables with respect to the local, separate universe cosmology with a subscript $W$, for ``windowed". The separate universe background matter density is defined as
		\begin{align}
			\bar{\rho}_{mW}(a) = \bar{\rho}_{m}(a)\lb 1 + \dl(a) \rb \, .
		\end{align}
		Here, the long-wavelength density contrast is defined $\dl = \delta \rho_m/\bar{\rho}_m$; the mean matter density in the global universe is $ \bar{\rho}_{m}$. We assume that the only contribution to the large-scale density perturbation is cold dark matter (CDM), but more general scenarios can be described by this formalism.  
	
		By absorbing $\dl$ into a local, mean matter density, an observer in this windowed region describes the universe according to a local Hubble rate $H_W$, scale factor $a_{W}$, and matter fraction $\om{}_{\! W}$. These cosmological parameters can be related to the global cosmology by requiring that the local, physical matter density matches the global one,
		\begin{align}
			\frac{\om H_0^2}{a^3}(1+\dl) = \frac{\om{}_{\! W} H_{0W}^2}{a_W^3} \, .
		\end{align}
		If $\dl$ is required to be a growing mode, the two cosmologies coincide at early times. To obtain an explicit transformation between the local and global cosmologies, we impose $a_W(a\rightarrow0) = a$, which implies $\om{}_W H_{0W}^2 = \om H_0^2$. Then, to linear order in the long-wavelength mode, the two cosmologies are related by 
		\begin{align}
			\label{eq:t1}
			a_W & \simeq a \lb 1- \frac{1}{3} \dl \rb  \, , \\
			\label{eq:t2}
			H_W & \simeq H \lb 1 - \frac{1}{3} \dl' \rb  \, , \\
			\label{eq:t3}
			\frac{d}{d\log \! a_W} & \simeq \lb 1 + \frac{1}{3}\dl' \rb \frac{d}{d\log\! a} \, .
		\end{align}
		Any small-scale observable $\mathcal{O}(a)$ has a counterpart in the windowed region, $\mathcal{O}_{W}(a\, |\, \dl)$, defined with respect to the separate universe cosmology in that region. The effect of the long-wavelength mode can be seen by expanding to linear order in $\dl$,
		\begin{align}
			\mathcal{O}_{W}(a\, |\, \dl) \simeq \mathcal{O}(a)\lb 1 + R_{\mathcal{O}}\dl \rb \, ,
		\end{align}
		where the linear, separate universe response is defined
		\begin{align}
			R_{\mathcal{O}} \equiv \frac{d\log\mathcal{O}_W}{d\dl}\bigg|_{\dl=0} \, .
		\end{align}
		In what follows we drop the notation indicating that the response is evaluated in the limit $\dl\rightarrow0$. In practice, we estimate the response by taking a finite difference derivative between observables in overdense and underdense boxes, so we only approximate this limit up to terms of order $\dl^{\ 2}$.
	
		One of the small-scale observables we study in this work is the cumulative number density of voids with effective radial size greater than some threshold, $n_{vW}(a, r > r_v \, |\, \dl)$. In this case, the separate universe response is the mean linear void bias
		\begin{align}
			\label{eq:br}
			b_v \equiv \frac{d\log n_{vW}}{d\dl} \, ,
		\end{align} 
		for voids larger than the threshold radius.
	
		The global background we use is the standard $\lcdm$ cosmology with Hubble rate given by
		\begin{align}
			\frac{H(a)}{H_0} = \sqrt{\frac{\om}{a^3} + \oL} \, ,
		\end{align}
		with cosmological parameters shown in Table~\ref{tab:cos}. The long-wavelength mode evolves according to the linear growth equation,
		\begin{align}
			\label{eq:dl}
			\dl'' + \lb 2 + \frac{H'}{H} \rb \dl' = \frac{3}{2}\frac{H_0^2}{H^2}\frac{\om}{a^3}\dl \, ,
		\end{align}
		where primes denote derivatives with respect to $\log (a)$. In this cosmology, long-wavelength perturbations undergo scale-invariant growth, which implies the bias in Eq.~(\ref{eq:br}) is scale invariant. In more general cosmologies, such as ones involving massive neutrinos with a free-streaming scale or clustering quintessence dark energy with a Jeans scale, the linear growth of long-wavelength perturbations is scale dependent, which leads to a scale-dependent bias. In this paper we focus exclusively on the scale-independent case.
	
	\section{Simulations}
		\label{sec:sim}

		\begin{table}
			\begin{tabular}{c c}
				\hline\hline
				Parameter & Value \\
				\hline
				$\oL$ & 0.7 \\			
				$\om$ & 0.3 \\
				$\ob$ & 0.05 \\
				$h$ & 0.7 \\
				$n_s$ & 0.968 \\
				$A_s$ & 2.137$\times10^{-9}$ \\ 
				$N_p$ & $(1024)^3$ \\
				$L_W$ & $1\ \rm{Gpc}/$$h$ \\
				$M_p$ & $1.108\times10^{11}\ \msun$\\
				\hline
				\hline
			\end{tabular}
			\caption{Cosmological and N-body simulation parameters.}
			\label{tab:cos}
		\end{table}

		\begin{figure*}
			\graphic{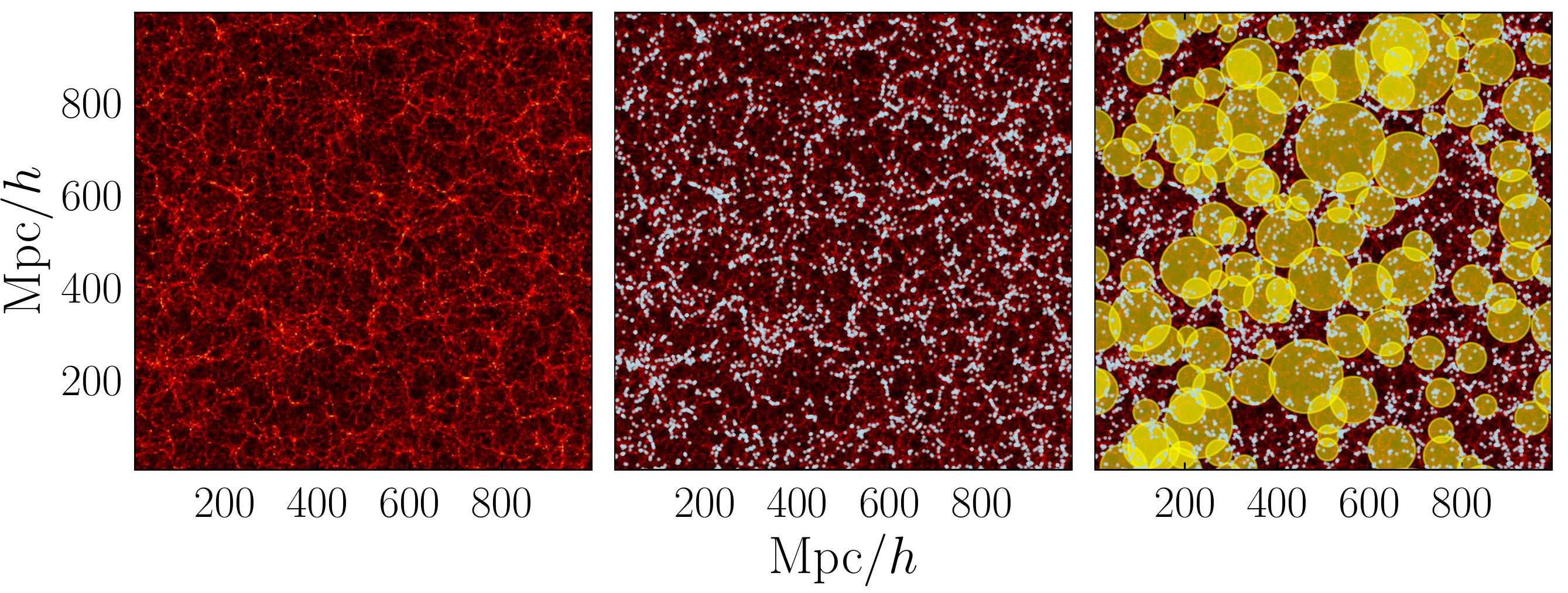}{1.0}
			\caption{Slices from simulation snapshots at redshift $z=0.00$, the slice has a thickness of $10$~Mpc/$h$. All plots show a log-scale heat map of the dark matter density field, the middle plot adds the halo positions for halos with mass $M>1.1\times 10^{13}\ \msun$, and the right plot adds the effective spherical area of the voids (identified in the halo distribution) that intersect the slice if their centers are no further than 1/3 their radial size outside of the slice. Note, this excludes some voids that do intersect with the slice, so some regions that appear underdense seem to contain no void.}
			\label{fig:sims}
		\end{figure*}

		We ran cosmological N-body simulations using the code Gadget-2~\cite{Springel:2005mi}. The code was modified to easily accommodate the separate universe scenario we studied. Rather than fixing cosmological parameters, from which Gadget-2 calculates the Hubble rate, a table of values was given relating the local Hubble rate to the local scale factor, $H_W(a_W)$. The modified code then interpolated values of the local Hubble rate at any given time step being evaluated. The local Hubble rate and local scale factor are calculated according to Eqs.~(\ref{eq:t1})~and~(\ref{eq:t2}). The long wavelength mode $\dl$ is fixed by integrating Eq.~(\ref{eq:dl}) with initial conditions fixed by require $\dl$ to be a growing mode with final values $\dl(a\rightarrow1) = \pm0.01$. The plus sign is for an overdensity and the minus sign is for an underdensity.  
	
		The simulations each contain $(1024)^3$ CDM particles in boxes with local, comoving length $L_W = 1\ \mathrm{Gpc}/h$. Note, since we fix the box size with respect to locally comoving coordinates, the underdense and overdense boxes have slightly different physical sizes. We chose this set up because it ensures that the dark matter particles have the same mass, $M_p = 1.108 \times 10^{11}\ \msun$, in both overdense and underdense boxes. 
	
		Random initial conditions for the simulations were generated by drawing random Fourier modes for the density field and velocity field. The random numbers were drawn from Gaussian distributions defined using a power spectrum, calculated with the program \textsc{class}~\cite{Blas:2011rf}. Although baryons behave as CDM at the scales of interest in our simulations, they affect the shape of the power spectrum through baryonic acoustic oscillations. To account for this, we calculate the power spectrum at the final time, $a=1$, with the parameters given in Table~\ref{tab:cos}. The spectral tilt and amplitude of the primordial power spectrum are taken from Particle Data Group's 2018 astrophysical constants and parameters list~\cite{Tanabashi:2018oca}. The power spectrum was rescaled back to the initial simulation time $a_i = 0.02$ using the local linear growth rate $D_W(a_W)$. Once the particle positions and velocities were calculated in real space by Fourier transforming the random fields, they were corrected using second order Lagrangian perturbation theory. This reduces transients in the initial conditions~\cite{Crocce:2006ve}.   

		We generated 40 random realizations of initial conditions for simulations in both the overdense and underdense boxes. The same random seeds were used to generate initial conditions for overdense and underdense pairs of simulations to maximize cosmic variance cancellation. In total, 80 simulations were run. Simulation snapshots were saved at redshifts $z=1.00$, $z=0.50$, and $z=0.00$. Note, the values of the simulation scale factor at which the snapshots are recorded are transformed according to Eq.~(\ref{eq:t1}) in order to match the simulation times with the previously listed redshifts in the global cosmology.  

		Simulation snapshots were analyzed using the halo finder Rockstar~\cite{Behroozi:2011ju}. This halo finder uses a friends-of-friends (FOF) algorithm to collect dark matter particles into halos and then analyzes their properties, such as mass, using spherical overdensity. This is important because, while the FOF linking length is difficult to map consistently between the global cosmology and the separate universe, the spherical overdensity thresholds are simple to transform. By identifying the halo masses using the spherical overdensity calculation, the resulting halo catalogs are not very sensitive to the FOF linking length. We modified Rockstar so that the spherical overdensity threshold was calculated with respect to the local, separate universe background.
	
		We also used the void finding program \textsc{vide}~\cite{Sutter:2014haa}, which utilizes the code \textsc{zobov}~\cite{Neyrinck:2007gy} to identify voids. \textsc{zobov} is based on a watershed algorithm that first tessellates the simulation box so that each tracer is associated with a cell containing space that is closer to it than any of the other tracers. This tessellation provides an estimation of the density field in which local density minima can be identified as the largest volume cells in a region. These density minima are joined with their surrounding cells into regions containing only one local density minima, which are called zones. In this way, the zones fully partition the volume of a simulation. Finally, voids are defined by joining zones together under the condition that the density ridge that separates them has at least one cell with density below $0.2$ times the background density. A slice from one of our simulation snapshots at redshift $z=0.0$ is shown in FIG. \ref{fig:sims}, including the positions of halos and voids.  
	
		Both the halo finder Rockstar and the void finder \textsc{vide} use density thresholds in their algorithms that need to be consistently mapped to the separate universe comoving coordinates. Rockstar uses the virial density from spherical collapse to define a halo's radial extent, and thus the amount of mass it encompasses. \textsc{vide} uses a density threshold (0.2 time the background density) in order to determine whether adjacent low density regions belong to the same void. In either case, these density thresholds are defined with respect to the global background cosmology and need to be altered in order to be consistent with the separate universe comoving coordinates. For a density threshold $\rho_{Th}$ in the global cosmology, the corresponding separate universe threshold is
		\begin{align}
			\rho_{ThW} = \rho_{Th}(1 - \dl)\, .
		\end{align}
		That is, an overdense region has a lower effective threshold, and an underdense region has a higher threshold. We modified both Rockstar and \textsc{vide} to ensure the appropriate density threshold was applied. In addition to this, we use a modification of the halo mass calculation in Rockstar, explained in \cite{Jamieson:2018biz}, which avoids the issue of having discrete mass values in multiples of the simulation particle mass. 
	
		We identified voids using two different catalogs of tracers. The first catalog included halos with mass greater than $1.1 \times 10^{13}\ \msun$. This corresponds to halos with 100 particles or more. The mean Eulerian bias of these halos is $b_h=1.46$ at redshift $z=0.0$, $b_h = 1.98$ at $z=0.5$, and $b_h = 2.73$  at $z=1.0$. The corresponding number densities are $2.0\times 10^{-4}$~Mpc$^{-3}$, $1.5\times 10^{-4}$~Mpc$^{-3}$, and $1.0\times 10^{-4}$~Mpc$^{-3}$ for those redshifts, respectively.  The results for these halo-traced voids are shown in Section \ref{sec:bias} and Section \ref{sec:prof}. The second catalog used in Section \ref{sec:dm} included dark matter particles of the N-body simulations, downsampled by a factor of 0.01. The number density of downsampled dark matter particles is $3.7\times 10^{-3} $~Mpc$^{-3}$ at all redshifts. The downsampling algorithm provided by \textsc{vide} randomly selects the particles, and we modified this to ensure the same set of particles is selected in overdense and underdense pairs with the same initial conditions. Otherwise, the random downsampling would slightly decreases cosmic variance cancellation for the voids bias.
	
	\section{Bias of voids found in halos}
		\label{sec:bias}
	
		\subsection{Definitions and Methods}

			The cumulative void size function, $n_v(r_{Th})$, is defined as the mean comoving number density of voids with effective radial size larger than $r_{Th}$. Expressed as an integral over the differential void size function it is given by
			\begin{align}
				\label{eq:n_v}
				n_v(r_{Th}) = \int_{r_{Th}}^{\infty} \frac{dr}{r} \lb -\frac{d n_v}{d\log r} \rb \, .
			\end{align}
			The minus sign accounts for the fact that the argument of $n_v$ is the lower bound of the integral. In separate universe regions, the void size function will be affected by the long wavelength mode in several ways. The comoving volume will be different, due to the difference in local scale factor, which changes the comoving number density of voids. Similarly, the effective radial size of the voids will be different for the same reason. These differences are trivial in the sense that they are a direct consequence of the coordinate change between the global and locally defined background cosmology, and running simulations is not required to determine these differences. The void size function will also differ due to the cumulative effects of the growth history of the long wavelength mode, which changes the local, separate universe expansion history. Determining this dynamical effect requires running simulations. We express the linear separate universe response of the void size function as a sum of these contributions,
			\begin{align}
				\frac{d\log n_{vW}(r_W|\dl)}{d\dl} = 1 + \frac{1}{3}\frac{d\log n_v}{d\log r} + \frac{\partial\log n_{vW}(r|\dl)}{\partial\dl} \, .
			\end{align}
			The first term on the right-hand side of the above expression accounts for the shift in comoving volume between the local and global cosmologies. The third term on the right-hand side is the Lagrangian linear void bias, and it is evaluated at equal comoving volume with respect to the separate universe scale factor. The second term, which is proportional to the differential void size function, can be removed if we first rescale the void radii to their corresponding value that is comoving with respect to the global cosmology. To be completely explicit, we remap the comoving radii measured in the separate universe $r_W$ to their corresponding values in the global universe $r$ via
			\begin{align}
				r =\frac{a_W}{a}r_W \simeq\lb1-\frac{1}{3}\dl\rb r_W\,.
			\end{align}
		
			Then, the remaining two terms sum to give the Eulerian linear void bias, which we refer to simply as the void bias,
			\begin{align}
				b_v & = 	\frac{d\log n_{vW}(r|\dl)}{d\dl} \\
					  & = 1 + \frac{\partial\log n_{vW}(r|\dl)}{\partial\dl} \, .
			\end{align}
			The bias can thus be determined by simply counting, or making a histogram, of voids in an overdense and underdense pair of separate universe simulations,
			\begin{align}
				b_v \simeq 1 + \frac{1}{\dl}\frac{n_{vW}(r|+\dl) - n_{vW}(r|-\dl)}{n_{vW}(r|+\dl) + n_{vW}(r|-\dl)} \, ,
			\end{align} 
			where all number densities are measured at a fixed, local, separate universe comoving volume. This method, however, is inefficient due to the scatter introduced at the sharp bin edges, which leads to statistical errors in the void bias greater than $10\%$.
	
			For a better determination of the bias, we used the abundance matching technique proposed by Li, Hu, and Takada~\cite{Li:2015jsz}. First, we combined all of the void catalogs from our separate universe simulation pairs into two cumulative void catalogs, one for the overdense simulations and one for the underdense simulations. Next, we rank ordered these cumulative catalogs from largest (rarest) to smallest (most abundant) voids, and truncated the longer list to match the length of the shorter one. Next, we construct the following lists:
			\begin{align}
				\log(r_i) & = \frac{\log(r_i^+) +\log(r_i^-)}{2} \, , \label{eq:amb1} \\
				n_{v i}  &= \lb i - \frac{1}{2} \rb \frac{1}{L_{W}^3 N_{\mathrm{sim}}} \, ,  \label{eq:amb2} \\
				s_i & =  \frac{\log(r_i^+) -\log(r_i^-)}{2\dl} \, .  \label{eq:amb3}
			\end{align}
			Here, the $\pm$ superscript refers to the overdense and underdense simulations. The first list estimates the mean radial size at fixed void abundance in the global universe. The second list estimates the cumulative size function in the global universe. The third list estimates the shift in the radial void size required to match the abundance of voids in the overdense and underdense simulation pairs. We then fit these lists with cubic splines to estimate the functions $n_v(r)$ and $s(r)$. We will refer to the latter function as the \emph{shift function}. The void bias is then given by
			\begin{align}
				b_v(r) = 1 - s(r)\frac{d\log n_v}{d\log r}\, .
			\end{align}
	
		\subsection{Simulation Results}
		
			\begin{figure*}
				\includegraphics[width=\linewidth]{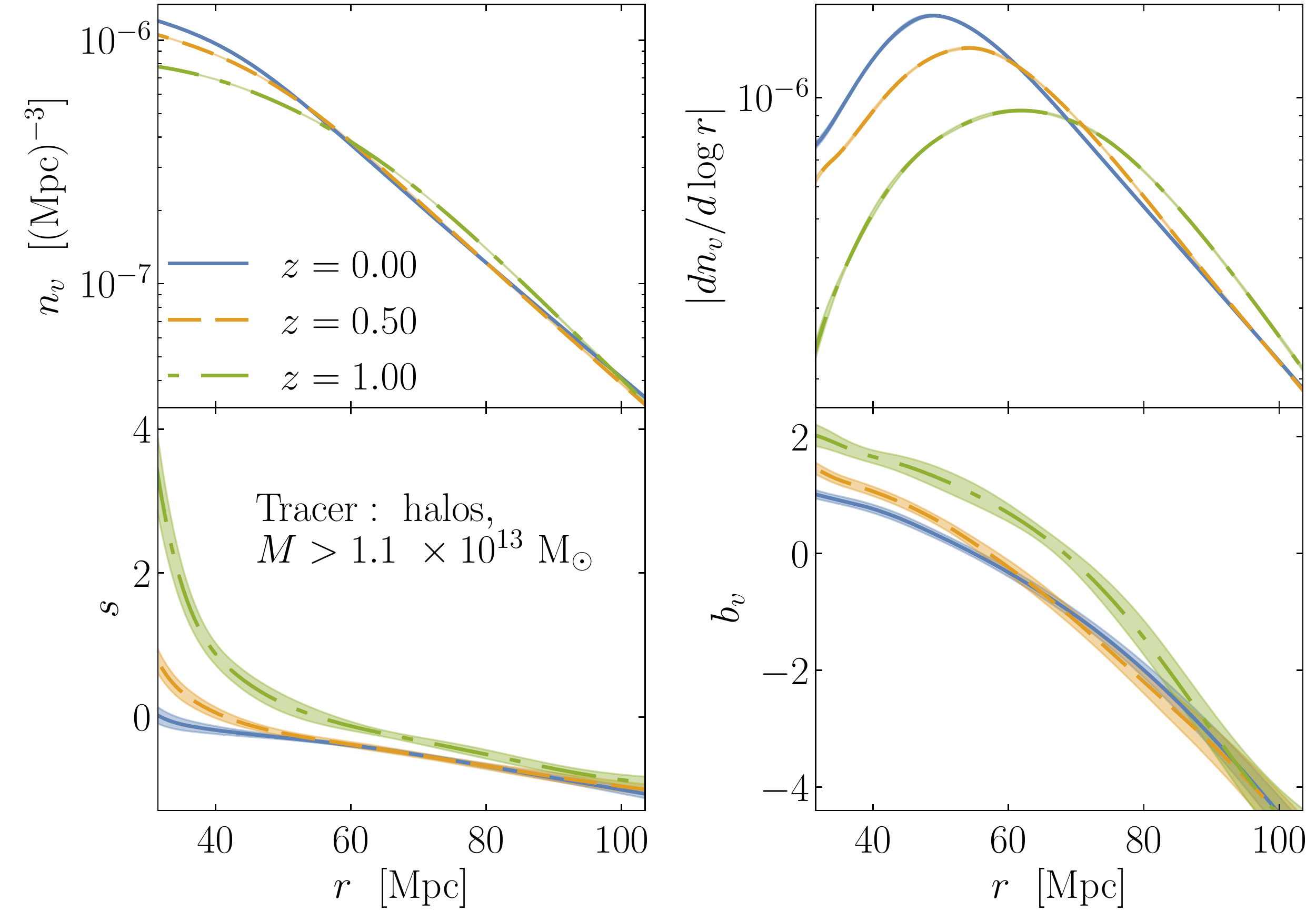}
				\caption{For all plots, the voids here are taken from the \textsc{vide} void finding algorithm, found in the distribution of simulation halos output by the halo finder Rockstar. The halo catalog includes all halos with mass greater than $1.1\times 10^{13}\ \msun$, which corresponds to a minimum of 100 particles per halo. Top left: cumulative void number density with effective radii greater than $r$. Top right: Differential void number density. Bottom Left: Shift in of the void size required to match void abundance between the overdense and underdense separate universe simulations. Bottom right: Eulerian void bias. The blue solid curves are from redshift $z=0.00$, the dashed orange curves are from redshift $z=0.50$, and the dash-dotted green curves are from redshift $z=1.00$. The shaded area around each curve indicates the 1-$\sigma$ bootstrap error.}
				\label{fig:br}
			\end{figure*}

			The void results from the abundance matching method are shown at three different redshifts, $z=0.0,\ 0.5,\ \mathrm{and}\ 1.0$, in FIG. \ref{fig:br}, along with the splines required to determine the bias. The shaded region indicates the bootstrap error over the 40 different random realizations of initial conditions for the simulation pairs. The mean separation of halos used to trace the voids at our highest redshift is roughly $15$~Mpc, and this is the smallest void size \textsc{vide} outputs. We show our results only for voids larger than $30$~Mpc but perform the spline fits over the smaller voids in the catalog to avoid edge effects. The bias is determined for voids of radial size up to about $100$~Mpc, above which the voids are extremely rare, and the abundance matching shift function is not well determined. Note, the spatial extent of these largest voids along one dimension is nearly $15\%$ of the length of the box sides.
	
			Qualitatively, the shift function at fixed redshift is monotonically decreasing with increasing void size and can be described as being positive for small voids and negative for large voids. The void size at which the shift function crosses zero evolves with redshift from larger to smaller voids. This leads to a void bias that is positive for small voids and negative for large voids, although the zero crossing of the Eulerian void bias is at larger void size than the zero crossing of the shift function. 
		
			The change in bias sign can be understood physically in terms of the environment in which voids of different sizes are found. A smaller void will typically be found in a region where the tracers are dense. For the halos we consider, the overdense boxes contain more halos but more importantly the halos are also more strongly clustered on small scales. We therefore expect more smaller voids in the overdense box, and consequently the small voids have positive bias. For larger voids, on the other hand, their existence requires broad depressions in density field of tracers. Since there are fewer halos in the underdense box, we expect larger empty regions, although we do not expect their abundances to be very different in the overdense and underdense boxes. Additionally, the underdense box undergoes more expansion, so the large voids, at fixed abundances, grow to be even larger in the underdense box than they are in the overdense box. This makes the shift function negative and leads to negative voids bias for large voids. This is consistent with the classification of small voids being considered overcompensated, in the sense that they occur in overdense environments, and large voids being undercompensated so that they occur in underdense environments \cite{Ceccarelli:2013rza}.  If the void is in an overdense environment the void-matter correlation is positive at $r\gtrsim r_v$ and the bias is positive, if the void is in an underdense environment the void-matter correlation function is negative at $r\gtrsim r_v$, and the bias is negative \cite{Sheth:2003py}.
	
			The redshift evolution of the void bias is mild for large voids. In fact, it is consistent with being constant for voids larger than $80$~Mpc between redshifts $z=1.0$ and $z=0.0$. Similarly, the bias is consistent with being constant between redshifts $z=0.5$ and $z=0.0$ for voids larger than $60$~Mpc. For voids smaller than $60$~Mpc, the bias decreases monotonically between redshifts $z=1.0$ and $z=0.0$. Small voids are rarer earlier on because they are positively correlated with clustered, collapsed objects, which are themselves rarer and less clustered at earlier redshifts. Larger voids are always rare, and negatively correlated with clustered, collapsed objects.  
	
		\subsection{Bias Decomposition}
		
			\begin{figure}
				\includegraphics[width=\linewidth]{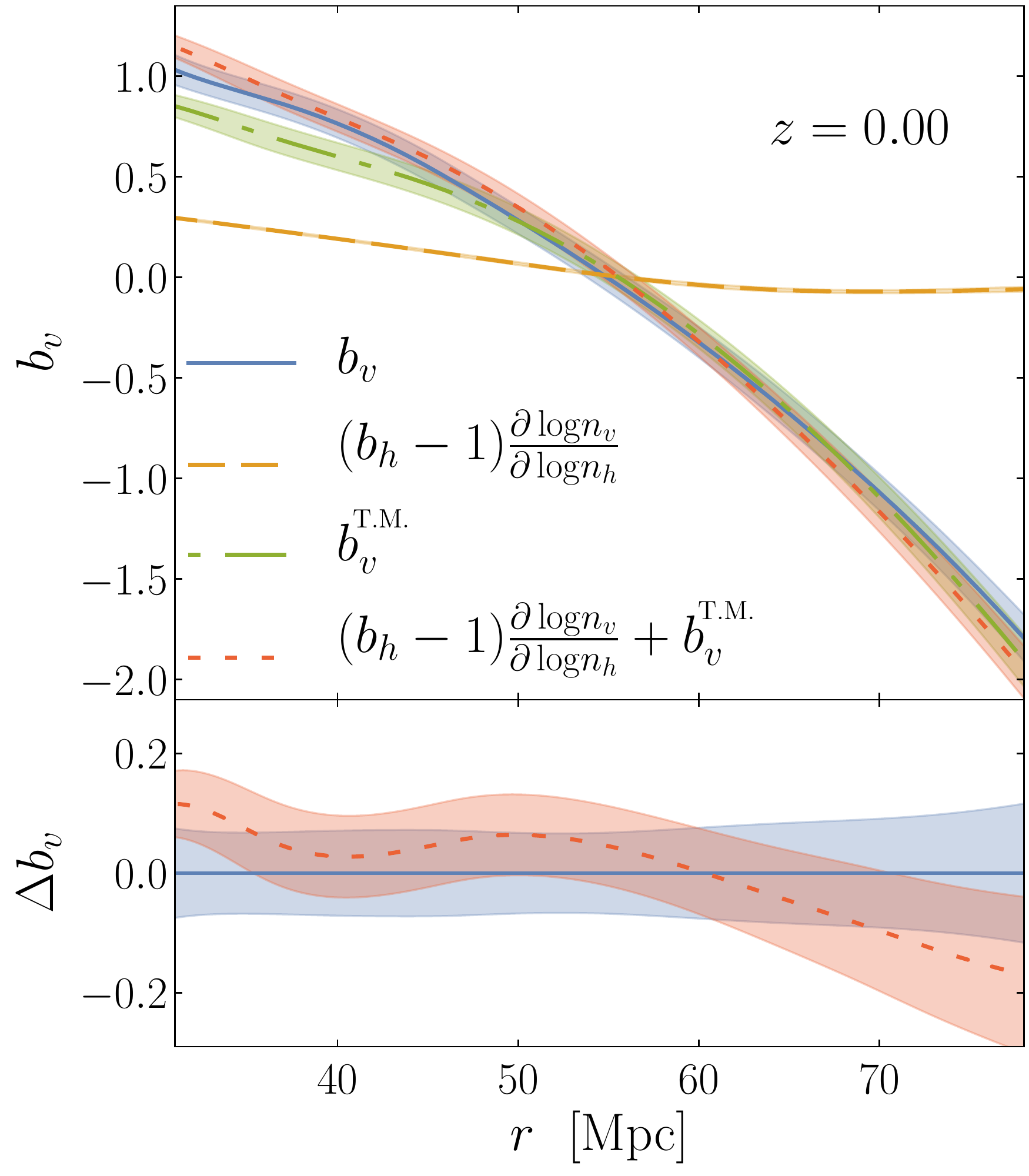}
				\caption{Void bias decomposition at redshift $z=0.0$ (see Eq.~(\ref{eq:bias_decomp})). The contribution from the shift in number density of tracers (halos) in over and underdense regions is shown by the dashed orange curve. The tracer-matched part of the bias, $b_v^{\!{}^{\mathrm{T.M.}}}$ is shown by the dot-dashed green curve. The solid blue curve shows the total bias while the dotted red curve shows the sum of the two contributions. The bottom panel shows the difference between the sum of the individual contributions and the total bias. The shading around each curve indicates the 1-$\sigma$ bootstrap error.}
				\label{fig:bdecomp}
			\end{figure}

			\label{ssec:decomp}
			The linear bias of voids found in biased tracers, such as halos, can be decomposed into two contributions. Part of the void bias is due simply to the fact that the mean number density of halos is different in overdense and underdense regions and the void population depends on the halos used to identify them. The remaining contribution to the void bias is due to changes in the gravitational clustering of tracers within overdense versus underdense regions. By including the number density of the tracers as a parameter on which the void size function depends, the void bias can be decomposed as
			\begin{align}
				\label{eq:bias_decomp}
				b_v = \lb b_h - 1\rb\frac{\partial\log n_{v}}{\partial\log n_h} +  \frac{\partial\log n_{vW}}{\partial \dl}\bigg|_{n_h} \, .
			\end{align}
			Note, the Lagrangian halo bias, $(b_h -1)$, appears in the above expression rather than the Eulerian bias because the effect of varying the halo density is measured at a fixed comoving box size, by varying the smallest mass allowed in the halo catalog. The first term can be measured in simulations, or even real data, by varying the number density of halos and finding the corresponding change in the void size function in the global background cosmology. The second term can be isolated in simulations by matching the abundance of tracers in an overdense and underdense separate universe simulation pair, and then finding the void bias. We define this to be the tracer-matched bias,
			\begin{align}
				b_v^{\!{}^{\mathrm{T.M.}}} = \frac{\partial\log n_{vW}}{\partial \dl}\bigg|_{n_h} \, .
			\end{align}
	
			The separate contributions to the void bias are plotted in FIG. \ref{fig:bdecomp}, along with their sum, which is compared with the full bias as determined from our simulations. The sum of the individual contributions is in agreement with the total bias within the 1-$\sigma$ bootstrap error. The amount of bias due to the change in tracer abundance is subdominant compared with the tracer-matched bias across the range of redshifts we consider $z=0-1$. As we expect, the effect of the number density of tracers is larger for small voids, which are found predominantly in regions of higher density. The shift in tracer density also has a small, negative effect on the bias for large voids. However, the total bias is indistinguishable from the tracer-matched bias for voids larger than $40$~Mpc. 
	
		\subsection{Validation Against Clustering Bias}
		
			\begin{figure*}
				\includegraphics[width=\linewidth]{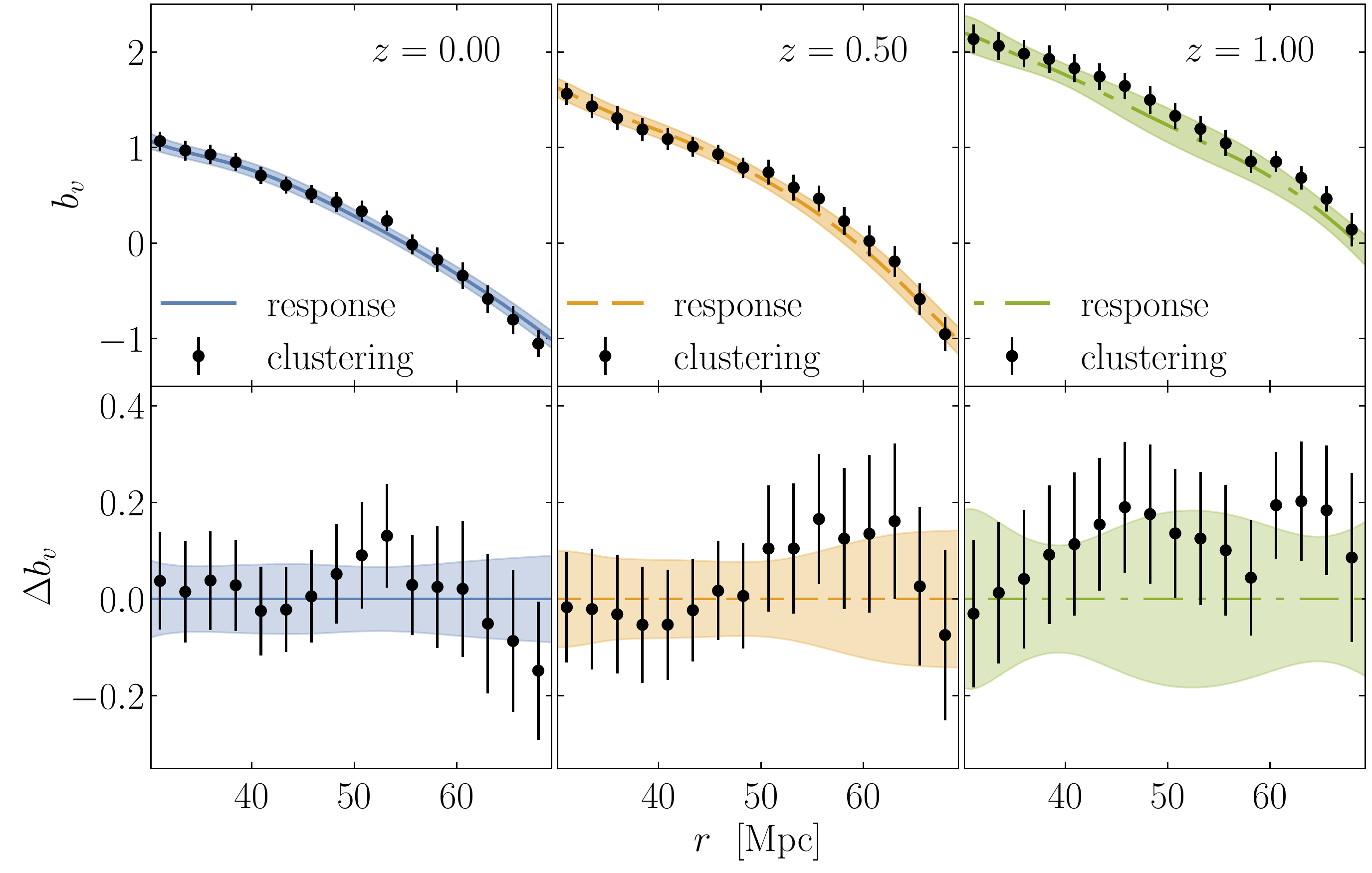}
				\caption{Top: Eulerian void bias for voids identified in halos at redshifts $z = 0.00$, $0.50$, and $1.00$, from left to right. The curves show the separate universe response bias with the shaded region indicating the 1-$\sigma$ bootstrap error. The black dots show the clustering bias with error bars corresponding to the 1-$\sigma$ bootstrap error. Bottom: difference between the clustering bias and the response bias.}
				\label{fig:hcomp}
			\end{figure*}
	
			\begin{figure*}
				\includegraphics[width=\linewidth]{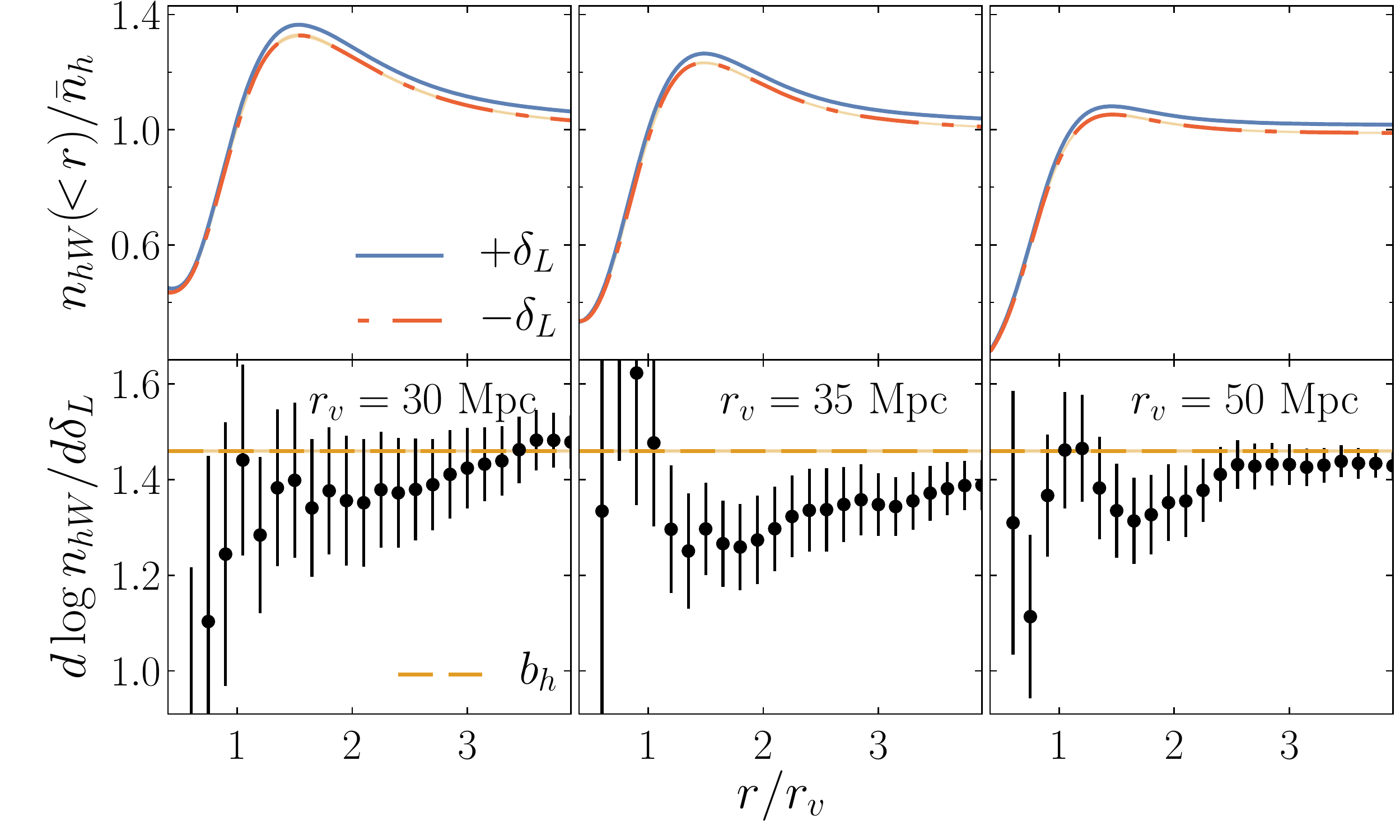}
				\caption{Top: void profiles as traced by the halos at redshift $z=0.0$, in the overdense (solid, blue), and underdense (dot-dashed, red) simulations. The profiles are defined with respect to the mean halo density in the global universe. The voids are stacked in bins, which from left to right are $27$~Mpc~$<r_v<33$~Mpc, $33$~Mpc~$<r_v<38$~Mpc, $47$~Mpc~$<r_v<54$~Mpc. The effective radii, $r_v$, give the mean radius of the stacked voids. Bottom: The separate universe response of the void profile at each radial bin. The dashed orange line indicates the mean halo bias for the halos used to trace the voids. At large radial distance, the void profile response should approach the halo bias, and does for all examples we test, though for the $r_v = 35$~Mpc example, this does not occur until $r/r_v\simeq6$.}
				\label{fig:vhprof}
			\end{figure*}

			In order to validate our determination of the void bias using separate universe methods, we compare our separate universe bias measurements to the clustering bias measured in the global cosmology. The clustering bias is defined as
			\begin{align}
				b_v^{c} = \lim_{k \rightarrow 0} \frac{P_{vm}(k)}{P_{mm}(k)} \, ,
			\end{align}
			where $P_{vm}$ is the void-matter cross power spectrum and $P_{mm}$ is the matter power spectrum. Since our simulations have a finite size, the fundamental mode in the box determines the lowest value of $k$ at which we can measure correlations. The fundamental mode in our simulations is $k_F = 0.0044~\mathrm{Mpc}^{-1}$. We bin the power spectrum in bins of width $k_F$, and use the points with $k<0.023~\mathrm{Mpc}^{-1}$ to determine the clustering bias.
	
			 We fit the bias to the form,
			\begin{align}
				\frac{P_{vm}(k)}{P_{mm}(k)} \simeq b_v^c + b_{k^2} k^2 + b_{k^4}k^4 \, ,
			\end{align}
			treating $b_{k^2}$, and $b_{k^4}$ as free parameters, and extract the clustering bias. The values of the clustering bias determined from this method are robust to our fit choices. A change in the maximum $k$-value by a factor of $2$ along with a change in the polynomial order to $k^2$ or $k^6$, changes the final values of $b_v^c$ by $\lesssim10\%$. 
	
			In general, the clustering bias for voids is less well-determined than measurements of the clustering bias for halos. Partly, this is due to the fact that there are fewer voids than halos; there is roughly 1 void for every 100 halos that we use as tracers. The large spatial extent of voids also makes it difficult to measure the clustering bias. On scales where $k \gtrsim r^{-1}$ , the ratio $P_{mv}/P_{mm}$ corresponds to the  Fourier transformation of the void profile. By analogy with the halo-model, this is the contribution from the 1-void term in the matter-void correlation function, whereas the clustering bias comes from the 2-void term. For voids with radial size greater than about $70$~Mpc, we are unable to distinguish the contributions from the 1-void and 2-void terms. Larger box sizes would be required to avoid having the void profiles affect scales where we perform our fits. Note that the separate universe-determined biases are not subject to this limitation.
		
			The comparison of the separate universe response bias and the clustering bias is shown in FIG. \ref{fig:hcomp} for redshifts $z=0.0$, $0.5$, and $1.0$. We find good agreement between the two biases for all redshifts. The statistical error of the separate universe response bias is slightly smaller than the clustering bias at redshifts $z=0.0$ and $z=0.5$. At void sizes away from where the bias crosses zero, the errors in the response bias range from $5\%\--10\%$, whereas the clustering bias has errors greater than $10\%$ at redshift $z=0.0$. 
	
			Finally, we note that the bias decomposition described in Section \ref{ssec:decomp} can also be performed in the global universe. For instance, the tracer-matched contribution to the bias can be isolated by combining measurements of the void and halo clustering biases from the halo-matter and void-matter cross power spectra as 
			\begin{align}
				\label{eq:clustering_decomp}
				b_v^{T.M.} &=\lim_{k\rightarrow 0}\left( \frac{P_{vm}(k)}{P_{mm}(k)} - \left(\frac{P_{hm}(k)}{P_{mm}(k)} -1\right)\frac{\partial \log n_v}{\partial \log n_h}\right)\\
				&=b_v^c - (b_h^c-1)\frac{\partial \log n_v}{\partial \log n_h}
			\end{align}
			where in the second line we have used $b_h^c = \lim_{k\rightarrow 0}P_{hm}(k)/P_{mm}(k)$ as the halo clustering bias. Since our clustering and response biases are in good agreement for both voids and halos, the decomposition determined from Eq.~(\ref{eq:clustering_decomp}) is trivially in agreement with FIG. \ref{fig:bdecomp} so we do not show it here. 
				
	\section{Profile Response for Voids Found in Halos}
		\label{sec:prof}
	
		The dynamics of void expansion produces a characteristic shape for the radial profile of voids. As the void expands, decreasing in density, it pushes matter outwards, which accumulates at the edge of the void in an overdense ridge \cite{Sheth:2003py}. The shape of the profile is affected both by expansion, in the void interior, and by clustering outside the void. Both of these processes respond to the presence of a long-wavelength perturbation. 
	
		We define the radial profile of a void as the mean number density of tracers within a radial distance $r$ of the voids center divided by the mean density in the global universe,
		\begin{align}
			\frac{n_{hW}(<r)}{\bar{n}_h} = \frac{3 N_{hW}(<r)}{4\pi r^3} \frac{L_W^3}{N_{hW, tot}} \lb 1 + \delta_L b_h   \rb \, .
		\end{align}  
		Here, we have defined $N_{hW}(<r)$ to be the sum of all halos in the separate universe simulation within a distance of $r$ for the void's center, and $N_{hW, tot}$ is the total number of halos in the box. The factor in the parentheses above converts the separate universe halo number density to the corresponding value in the global universe, $\delta_L$ is positive for an overdense box and negative for an underdense box. Notice, since we define the void profiles with respect to the global halo density, the profile asymptotically approaches
		\begin{align}
			\lim_{r\rightarrow\infty}\frac{n_{hW}(<r)}{\bar{n}_h} =\lb 1 + \delta_L b_h   \rb \, ,	
		\end{align}
		and the asymptotic separate universe response of the profile, by definition, is
		\begin{align}
			\lim_{r\rightarrow\infty}\frac{d\log n_{hW}(<r)}{d\delta_L} = b_h \, .	
		\end{align}
	
		We measured the void profiles at 40 radial positions out to 6 times the void radius. The profiles  were stacked as a function of $r/r_v$, where $r_v$ is the radius of an individual void.  Profiles are shown for a representative set of voids binned based on radial size, along with the separate universe response, in FIG. \ref{fig:vhprof}. The bins include voids of effective radii  $27$~Mpc~$<r_v<33$~Mpc, $33$~Mpc~$<r_v<38$~Mpc, $47$~Mpc~$<r_v<54$~Mpc.
		
		Asymptomatically, the profile responses measured in our simulations do approach the halo bias. However, the rate at which the profiles converge to their asymptotic value depends on void size. For smaller voids ($r_v \simeq 30$~Mpc), the profile response converges to the halo bias at a radial distance corresponding to $r/r_v \simeq 3$. For larger voids ($r_v \simeq 50$~Mpc), the profile response converges closers to $r/r_v\simeq 2$, which is a somewhat larger physical distance from the void center ($r\simeq90$~Mpc for the small voids, $r\simeq100$~Mpc for the large voids). For voids of intermediate size  $r_v \simeq 35$~Mpc, the convergence is slower. While the response is within 10\% of the bias at $r/r_v=3$, it does not approach percent level agreement with the halo bias until $r/r_v\simeq6$, or $r\simeq210$~Mpc.
	
		The overdensity at the void edge typically spans the radial range $1<r/r_v<3$. In this region, we find the profile response is below the halo bias. This suggests the halos that populate the void edge are less clustered than the mean halo in the box. That is, the number density of halos occupying the void edge is less sensitive to the long-wavelength mode than the number density of halos in the whole box, and so these halos are less biased. The depth of the dip in the profile response is dependent on void size, and we find a maximum depression in the profile response for the voids of size $r_v\simeq35$~Mpc. At the edge of the void $r/r_v=1$, we find the profile response is consistent with the halo bias over the full range of void sizes from our simulations.
	
		We do not resolve the interior of the voids ($r/r_v < 1$) very well due to the sparseness of the halos that we used as tracers, so we cannot conclusively determine how sensitive the void interior is to the background density. However, the measured responses suggest the void interiors are not very sensitive to the long wavelength mode, and profiles from overdense and underdense simulations might become indistinguishable towards the void centers. As we shall see, the profiles of the voids found in downsampled dark matter have better resolution in the void interiors and they appear to be insensitive to shifts in the background density. 

	\section{Voids found in downsampled dark matter}
		\label{sec:dm}
		
				\begin{figure*}
			\includegraphics[width=0.9\linewidth]{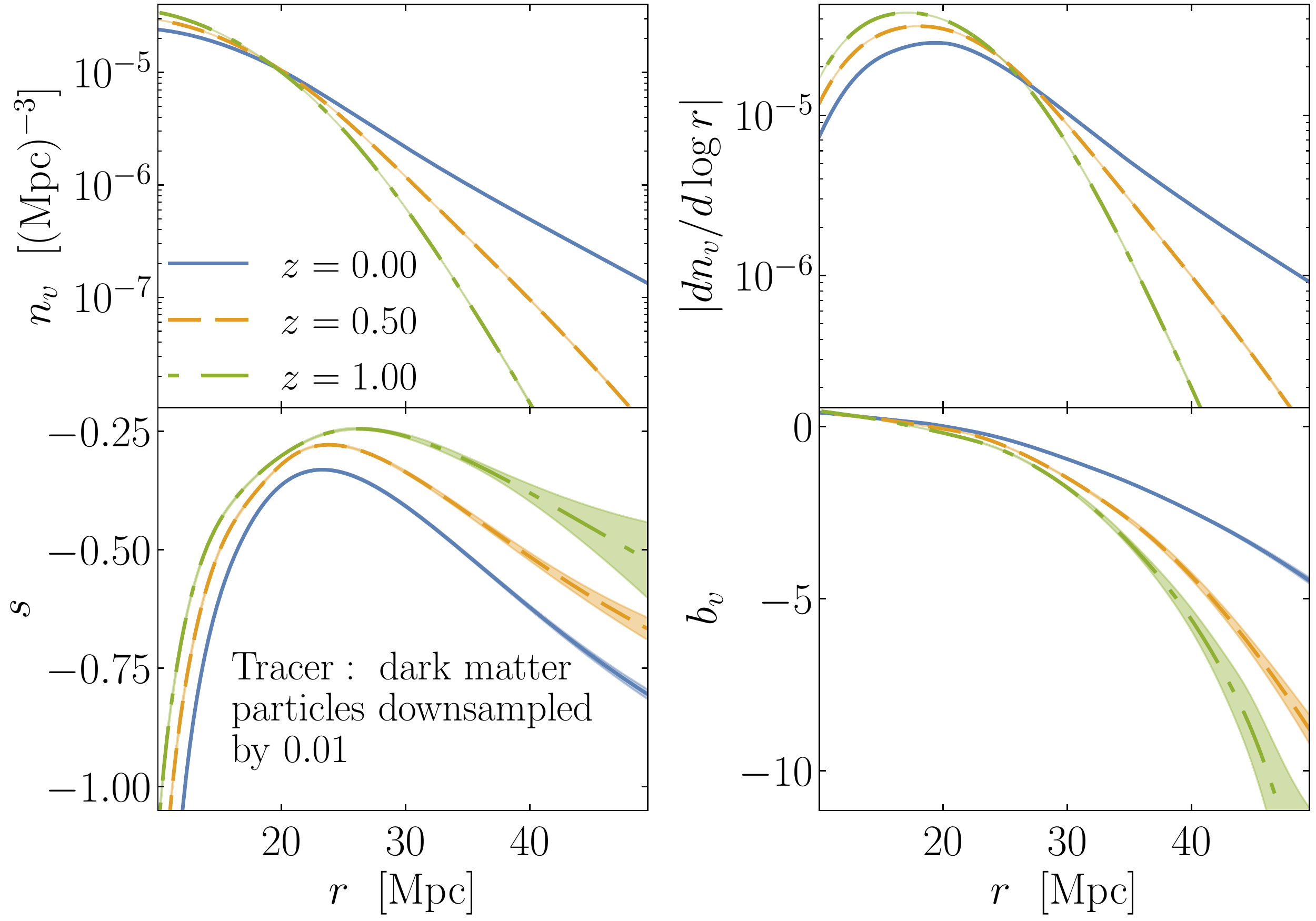}
			\vspace{-0.2in}
			\caption{For voids found in simulation dark matter particles, downsampled by a factor of 0.01. Top left: cumulative void number density. Top right: differential void number density. Bottom Left: void shift function. Bottom right: Eulerian void bias. The blue solid curves are from redshift $z=0.00$, the dashed orange curves are from redshift $z=0.50$, and the dash-dotted green curves are from redshift $z=1.00$. The shaded area around each curve indicates the 1-$\sigma$ bootstrap error.}
			\label{fig:dmbr}
			
			\vspace{0.1in}
			
			\includegraphics[width=0.9\linewidth]{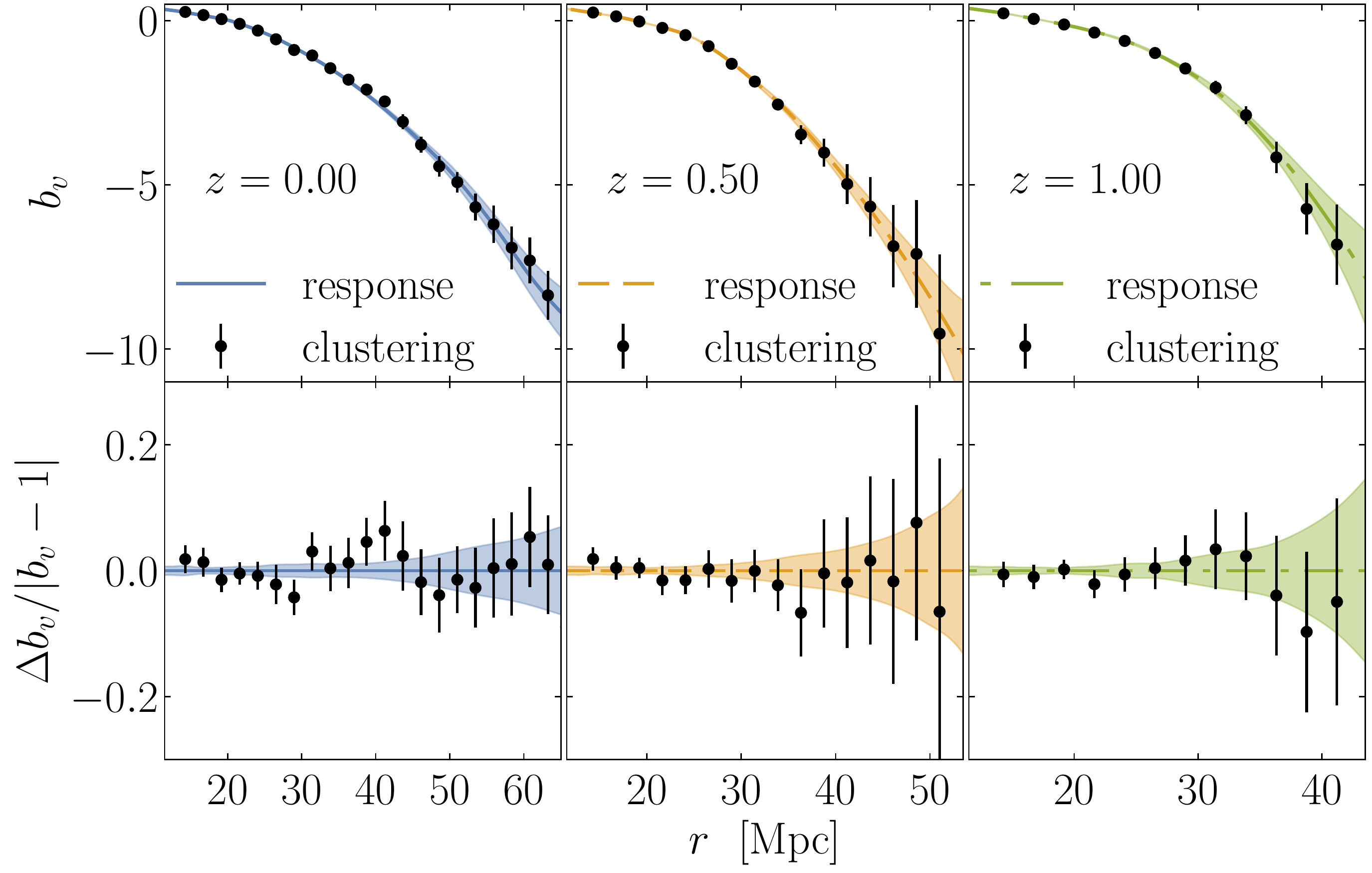}
			\vspace{-0.2in}
			\caption{Top: Eulerian void bias for voids found in dark matter, downsampled by a factor of 0.01, at redshifts $z = 0.00$, $0.50$, and $1.00$, from left to right. The curves show the separate universe response bias with the shaded region indicating the 1-$\sigma$ bootstrap error. The black dots show the clustering bias with error bars corresponding to the 1-$\sigma$ bootstrap error. Bottom: fractional difference between the clustering bias and the response bias relative to the Lagrangian void bias ($b_v -1$).}
			\label{fig:dmcomp}
		\end{figure*}
		
		\begin{figure*}		
			\includegraphics[width=\linewidth]{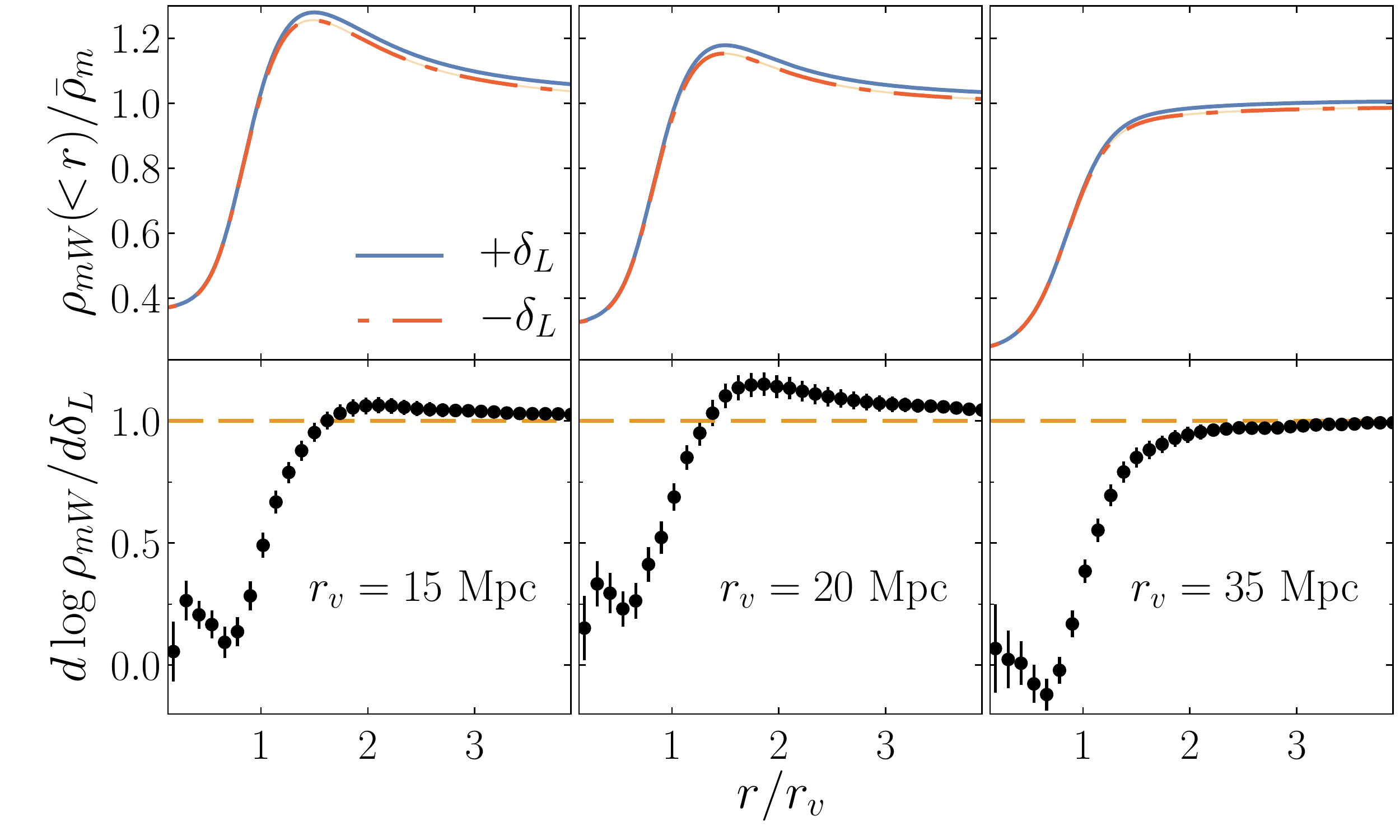}
			\caption{Top: void profiles as traced by the dark matter, for voids found in dark matter downsampled by a factor of 0.01 at redshift $z=0.0$, in the overdense (solid, blue), and underdense (dot-dashed, red) simulations. The profiles are defined with respect to the mean downsampled dark matter density in the global universe. The voids are stacked in bins, which from left to right are $14$~Mpc~$<r_v<16$~Mpc, $19$~Mpc~$<r_v<21$~Mpc, $33$~Mpc~$<r_v<37$~Mpc. The effective radii, $r_v$, give the mean radius of the stacked voids. Bottom: The separate universe response of the void profiles at each radial bin. The dashed orange line indicates the response of the background, separate universe matter density. At large radial distance, the void profile response should approach unity.}
			\label{fig:vdmprof}
		\end{figure*}
		
		In this section we present the linear void bias and the void profile response to long wavelength perturbations for voids found in the dark matter particles, downsampled by a factor of 0.01. For this analysis, we use the same set of simulations used for voids found in halos. Since we use the same initial conditions for the overdense and underdense pairs of simulations, we match the randomly downsampled particles in the pairs too, by identifying their particle ID numbers from Gadget2. This avoids introducing extra variance that would otherwise occur due to different particles being chosen when downsampling in the over and underdense boxes. 

		The abundance-matched spline method (see Eqs. [\ref{eq:amb1}--\ref{eq:amb3}]) was used to obtain the void size function, the shift function, and the bias. Our results for these quantities are shown in FIG. \ref{fig:dmbr}. Whereas the shift function for voids found in halos increases rapidly near the minimum void size for founds found in halos, the trend is opposite for voids in dark matter. The shift function is always negative and decreases rapidly to more negative values approaching the minimum void size. The shift function at fixed void size decreases monotonically with redshift for the voids in dark matter.  The Eulerian void bias has almost no change between redshifts $z=0.0$ and $z=1.0$ for small voids, near $10$~Mpc in radial size. These small voids have a bias of unity, although there is potentially a crossover of the bias curves from different redshifts that we do not have the sensitivity to measure. For larger voids, the bias increases monotonically with decreasing redshift.

		To validate our separate universe measurements, we determine the clustering bias by fitting,
		\begin{align}
			\frac{P_{mv}(k)}{P_{mm}(k)} = b_v^c + b_{k^2}k^2 \, ,
		\end{align}
		treating $b_{k^2}$ as a free parameter. For the voids in dark matter, this ratio of power spectra is flatter in the low $k$ limit than for halos, so we use a polynomial that is linear in $k^2$ and perform our fits for $k < 0.023$~Mpc$^{-1}$. 

		The comparison with the clustering bias is shown in FIG. \ref{fig:dmcomp}. The top panels show the Eulerian void bias, while the bottom panels show the fractional difference between the Lagrangian response bias and the Lagrangian clustering bias. We use the Lagrangian bias for the comparison because, unlike voids found in halos, the Lagrangian bias does not cross zero for the dark matter voids over the range of void sizes in our catalogs. We find good agreement between the clustering bias and the separate universe response bias measurements. The separate universe determinations of the void bias have a much smaller variance than the clustering bias measurements. For voids between radial size $20$~Mpc and $40$~Mpc the bootstrap errors in the response bias is on the order of 1\%. 
	
		The void bias for voids found in the halo field is not easily comparable to the bias of voids found in the downsampled dark matter field. The number densities of the two tracers are different orders of magnitude, and while the number density of halos increases with time, the number of dark matter simulation particles remains fixed. The halos are also biased tracers, while the dark matter particles are unbiased. The void bias for dark matter voids does not admit a decomposition like the one presented in Section \ref{ssec:decomp}, since this bias is purely tracer-matched already.

		Void profiles in dark matter and the sensitivity of the profiles to the shift in the background density are shown in FIG. \ref{fig:vdmprof}. Note that these profile responses asymptotically approach unity, corresponding to the shift in separate universe comoving volume. Unlike the halo profiles of voids found in halos, we find an excess of sensitivity in the profile in the region between 1 and 2 times the void's radial size for small voids. We have much better resolution for the profiles of voids found in dark matter, so we are able to determine the sensitivity of the void interiors.  We find that the response of the profile to the long-wavelength mode decreases from the void edge to the void center, suggesting that the profile in the interior of the void is nearly insensitive to the presence of the long-wavelength mode.
	
	\section{Conclusion}
		\label{sec:con}
		Cosmic voids are a promising new observable for studying cosmology through the late-time, large-scale structure of the Universe. The structure, evolution, and statics of voids, however, are far less understood than halos. In this paper, we have advanced these efforts by demonstrating the first measurement of linear void bias using separate universe simulations. The separate universe response bias measurements presented in Section \ref{sec:bias} and \ref{sec:dm} were shown to be in excellent agreement with measurements of the clustering bias for voids (FIG. \ref{fig:hcomp} and FIG. \ref{fig:dmcomp}). We have also shown that, for voids found in halos, the void bias can be decomposed into contributions coming from the bias of the underlying tracer, and dynamical contributions due directly to differences in local expansion histories arising from the presence of long-wavelength modes (FIG. \ref{fig:bdecomp}). The latter contribution dominates over the range of void sizes found in our simulations and across $z=0$ to $z=1$. 
	
		In addition to void bias, we also measured void profiles in our overdense and underdense separate universe simulations. The separate universe responses of the void profiles, presented in Section \ref{sec:prof} for voids identified in halos and in Section \ref{sec:dm} for voids identified in dark matter, indicate the sensitivity of void profiles to shifts in the background density. The responses of void profiles in halos and matter can also be considered one piece of the void-halo-matter and void-matter-matter three-point functions, respectively. The shapes of void profiles are most sensitive to shifts in the background density of the Universe at distances between $1$ to $3$ times a void's effective radial size, while the void interiors are much less sensitive. The sensitivity of a void's profile depends on the void's size. For the void and halo samples we considered, the void profile in halos is most sensitive to the background density for voids with effective radial size $r_v\simeq35$~Mpc. 
		\\ \indent
		This first demonstration and validation of separate universe methods applied to cosmic voids opens new possibilities for studying beyond $\Lambda$CDM physics using separate universe simulations. Analogously to halo bias, the linear void bias will be scale-dependent in the presence of long-wavelength modes that undergo scale-dependent growth, such as in the cases of massive neutrinos, and clustering quintessence. By extending the work we present here to these cosmologies, the sensitivity of cosmic voids to new physics can be determined using separate universe techniques. 

		\acknowledgements
		DJ and ML are grateful for helpful correspondence and conversations with Guilhem Lavaux, Federico Marulli, Ravi Sheth, and Ben Wandelt. Results in this paper were obtained using the high-performance computing system at the Institute for Advanced Computational Science at Stony Brook University. DJ is supported by grant NSF PHY-1620628 and DOE DE-SC0017848. ML is supported by DOE DE-SC0017848. \\~\\
		{\em {\bf Note added: }}Recently, another paper studying voids in the separate universe framework appeared on the arXiv \cite{Chan:2019yzq}. That paper determines the linear bias (as we do) and quadratic bias (which we do not study) for voids found in dark matter and halos. That paper does not study the bias decomposition nor the void profiles, which are included here. 

	\bibliography{void_paper.bib}
	
\end{document}